\documentclass[sigplan,nonacm]{acmart}

\usepackage{amsmath,amsfonts}
\usepackage{algorithmic}
\usepackage{graphicx}
\usepackage{textcomp}
\usepackage{xcolor}
\usepackage{booktabs} 
\usepackage{tikz}
\usepackage{makecell} 
\newcommand{\circled}[2][]{%
  \tikz[baseline=(char.base)]{
    \node[shape=circle, draw, fill=#1, minimum size=0.8em, inner sep=0pt, font=\footnotesize] (char) {\textcolor{white}{#2}};
  }%
}

\begin{document}

\pdfpagewidth=8.5in
\pdfpageheight=11in

\title{A Hybrid GPU and Processing-Near-Memory Architecture with Shared DRAM for Efficient Prefill–Decode Disaggregation}



\author{Sanghyeok Han}
\authornote{These authors contributed equally to this work.}
\email{hansh778@snu.ac.kr}
\affiliation{
  \institution{Seoul National University}
  \city{Seoul}
  \country{Republic of Korea}
}

\author{Inseong Choi}
\authornotemark[1]
\email{inseong.choi@snu.ac.kr}
\affiliation{
  \institution{Seoul National University}
  \city{Seoul}
  \country{Republic of Korea}
}

\author{Byungkuk Yoon}
\email{bkyoon@snu.ac.kr}
\affiliation{
  \institution{Seoul National University}
  \city{Seoul}
  \country{Republic of Korea}
}

\author{Gyeonghwan Park}
\email{gyeonghwan.park@snu.ac.kr}
\affiliation{
  \institution{Seoul National University}
  \city{Seoul}
  \country{Republic of Korea}
}

\author{Jae-Joon Kim}
\authornote{Corresponding author.}
\email{kimjaejoon@snu.ac.kr} 
\affiliation{
  \institution{Seoul National University}
  \city{Seoul}
  \country{Republic of Korea}
}

\begin{abstract}

The growing adoption of large language models (LLMs) has significantly increased the compute and memory demands of inference serving. 
A key challenge arises from the mismatch between compute-bound prefill and memory-bound decoding: when executed together, the two phases interfere, while separating them via prefill–decode (PD) disaggregation leads to substantial resource underutilization. 
Specifically, prefill leaves memory bandwidth underutilized, whereas decoding underutilizes compute resources.
Prior approaches, such as GPU Streaming Multiprocessor (SM) partitioning, attempt to improve utilization by spatially dividing GPU resources. 
However, these approaches remain fundamentally limited, as memory-bound decoding still requires a large number of SMs to saturate bandwidth, reducing the compute resources available for prefill and resulting in an inherent trade-off.
To address this limitation, we propose \textbf{DuoPD}, a hybrid PD-disaggregation architecture that integrates lightweight Processing-Near-Memory (PNM) units into the HBM logic die. 
Unlike prior PIM/PNM designs that rely on costly internal bandwidth scaling and offload only partial decoding operations, DuoPD enables full decoding execution on PNM by leveraging bandwidth sharing with the GPU over a shared HBM interface. 
This allows compute-bound prefill and memory-bound decoding to execute concurrently with minimal interference, improving both compute and memory utilization.
To realize this design, we develop a compact PNM microarchitecture, a TSV-level bandwidth sharing mechanism, and a hierarchical data mapping scheme tailored for hybrid GPU–PNM execution. 
Evaluations on server-scale LLMs demonstrate that DuoPD achieves up to 2.23$\times$ higher end-to-end throughput than conventional PD-disaggregation.

\end{abstract}

\maketitle

\vspace{-0.05in}
\section{Introduction}

The demand for efficient Large Language Model (LLM) \cite{GPT,LLaMA2-70B,palm} inference has surged with the rapid growth of applications such as chatbots and machine translations~\cite{chatbot, translation}. Due to their massive parameter sizes, LLMs require substantial compute and memory resources. To meet these demands, modern GPUs~\cite{H100,GB300} integrate a large number of compute cores and high-bandwidth memory (HBM) stacks~\cite{HBM3,HBM4}, but this trend significantly increases data center cost and power consumption.
To improve throughput and reduce serving cost, batching techniques process multiple user requests together~\cite{orca,multibatch,vLLM}. 
However, when prefill, which processes long contexts at once and decoding, which processes one token at a time, are executed within the same batch, decoding latency becomes bound by the longer prefill latency, leading to excessive tail latency in decoding.

\begin{figure}[t!]
\centerline{\includegraphics[width=\columnwidth]{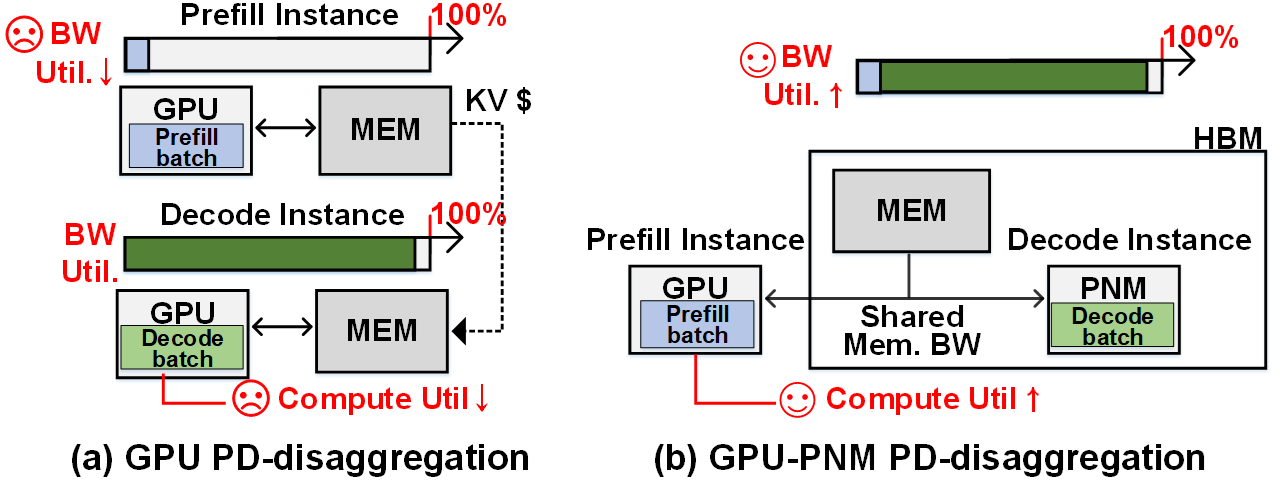}\vspace{-0.1in}}

\caption{Comparison of PD-disaggregation systems: (a) Conventional and (b) our GPU-PNM system (DuoPD).}
\label{fig:intro}
\vspace{-0.2in}
\end{figure}

Prefill–Decode (PD) disaggregation~\cite{distserve,windserve,splitwise,dynamo} addresses this problem by executing the two phases on separate devices (Fig.~\ref{fig:intro}(a)). This eliminates cross-phase interference and allows prefill and decoding requests to be batched independently. However, PD-disaggregation suffers from a major inefficiency: prefill workloads are compute-bound and underutilize memory bandwidth, while decoding workloads are memory-bound and underutilize GPU compute resources. As both compute and memory dominate system cost, this asymmetry leads to significant resource inefficiency.

Recent works attempt to mitigate this inefficiency via 
Streaming Multiprocessor (SM) partitioning~\cite{bullet,muxwise}, which spatially divides GPU resources between prefill and decoding. While this improves utilization compared to conventional PD-disaggregation, it fails to fully resolve the imbalance. In particular, even the memory-bound decoding requires relatively large number of SMs to fully utilize the DRAM bandwidth due to the complex cache and NoC architectures of GPU. As a result, substantial number of SMs must still be allocated to decoding despite the use of Tensor Memory Accelerator (TMA), leading to persistent contention with prefill and limiting overall efficiency, especially in modern GPUs with extremely high HBM bandwidth.

In this work, we present \textbf{DuoPD}, a hybrid GPU--Processing-Near-Memory (PNM) architecture which highly improves utilization of both compute and memory bandwidth (Fig.~\ref{fig:intro}(b)). DuoPD is based on two key observations. First, prefill inherently cannot maintain peak bandwidth of HBM due to its high arithmetic intensity, leaving substantial bandwidth consistently underutilized. Second, decoding's low arithmetic intensity underutilizes GPU compute resources, making it well suited for lightweight near-memory compute units, which can directly exploit HBM bandwidth without requiring large compute resources.

Leveraging these insights, DuoPD offloads the entire decoding execution to lightweight PNM units integrated within the HBM logic die, while dedicating all GPU compute resources to prefill. 
In DuoPD, GPU prefill and PNM decoding execute concurrently over a shared HBM interface. Specifically, their memory accesses are interleaved at the TSV interface: while the GPU consumes previously fetched data to perform compute-intensive prefill, the PNM simultaneously streams data from HBM at high bandwidth to serve memory-bound decoding. This complementary overlap enables DuoPD to reclaim otherwise idle memory bandwidth without requiring additional hardware bandwidth.

DuoPD differs fundamentally from prior PIM/PNM-based approaches in two key aspects. First, prior works primarily focus on increasing internal memory bandwidth to improve parallelism, often incurring significant area, power, and capacity overheads. In contrast, DuoPD improves \textit{effective bandwidth utilization} by exploiting underutilized bandwidth already available in existing systems, without requiring additional bandwidth provisioning. Second, prior approaches typically adopt \textit{intra-decode disaggregation}, where memory-intensive components such as attention are offloaded to PIM/PNM while the remaining decode operations are executed on GPUs within a conventional PD-disaggregation framework. In contrast, DuoPD performs \textit{inter-phase disaggregation} within a single device, executing prefill on the GPU and the entire decoding phase on PNM over shared memory. This unified execution model 
enables tightly coupled compute--memory co-execution.

To realize this design, we develop a compact PNM microarchitecture that fits within the tight area and power constraints of the HBM logic die while maintaining full memory bandwidth utilization. We further introduce two key mechanisms to enable efficient GPU--PNM coexistence. First, we design a direct data forwarding path from TSVs to the PNM buffer, bypassing the memory controller to eliminate redundant data movement and 
I/O contention. Second, we propose a memory management unit (MMU) that arbitrates GPU and PNM memory requests over the shared datapath, minimizing interference between prefill and decoding.
In addition, we propose a hierarchical data mapping scheme that efficiently distributes model parameters and KV caches across multiple PNMs while preserving compatibility with GPU execution.

Our contributions are summarized as follows:
\begin{itemize}
\item We identify \textbf{structural resource underutilization} as a fundamental limitation of PD-disaggregated LLM serving and show that it cannot be resolved by existing GPU-only approaches.
\item We present \textbf{DuoPD}, a hybrid GPU--PNM architecture that enables concurrent compute-bound prefill and memory-bound decoding over shared HBM, improving both memory/compute utilization.
\item We demonstrate that improving \textbf{effective memory bandwidth utilization}, rather than scaling physical bandwidth, achieves higher performance without incurring the area, power, and capacity overheads of prior PIM/PNM designs.
\item We design a lightweight PNM microarchitecture, a TSV-level data forwarding mechanism, and an MMU-based arbitration scheme to enable efficient GPU--PNM co-execution.
\item We show the simulation results that DuoPD achieves up to 2.23$\times$ higher end-to-end throughput and lower latency than GPU-based PD-disaggregation under the same GPU budget.
\end{itemize}

\vspace{-0.05in}
\section{Background}
\subsection{Interference between Prefill and Decoding}

LLM inference is divided into two distinct phases: prefill and decoding. 
In the prefill phase, the model processes the entire input sequence to construct the KV cache, and is dominated by compute-intensive GEMM operations. As long-context workloads such as retrieval-augmented generation (RAG)~\cite{rag} continue to grow, prefill becomes increasingly compute-bound.
In contrast, decoding generates one token at a time in an autoregressive manner. It mainly involves compute-light GEMV operations but requires frequent access to the KV cache, making it strongly memory-bound.


\begin{figure}[t]
\centering
\includegraphics[width=0.9\columnwidth]{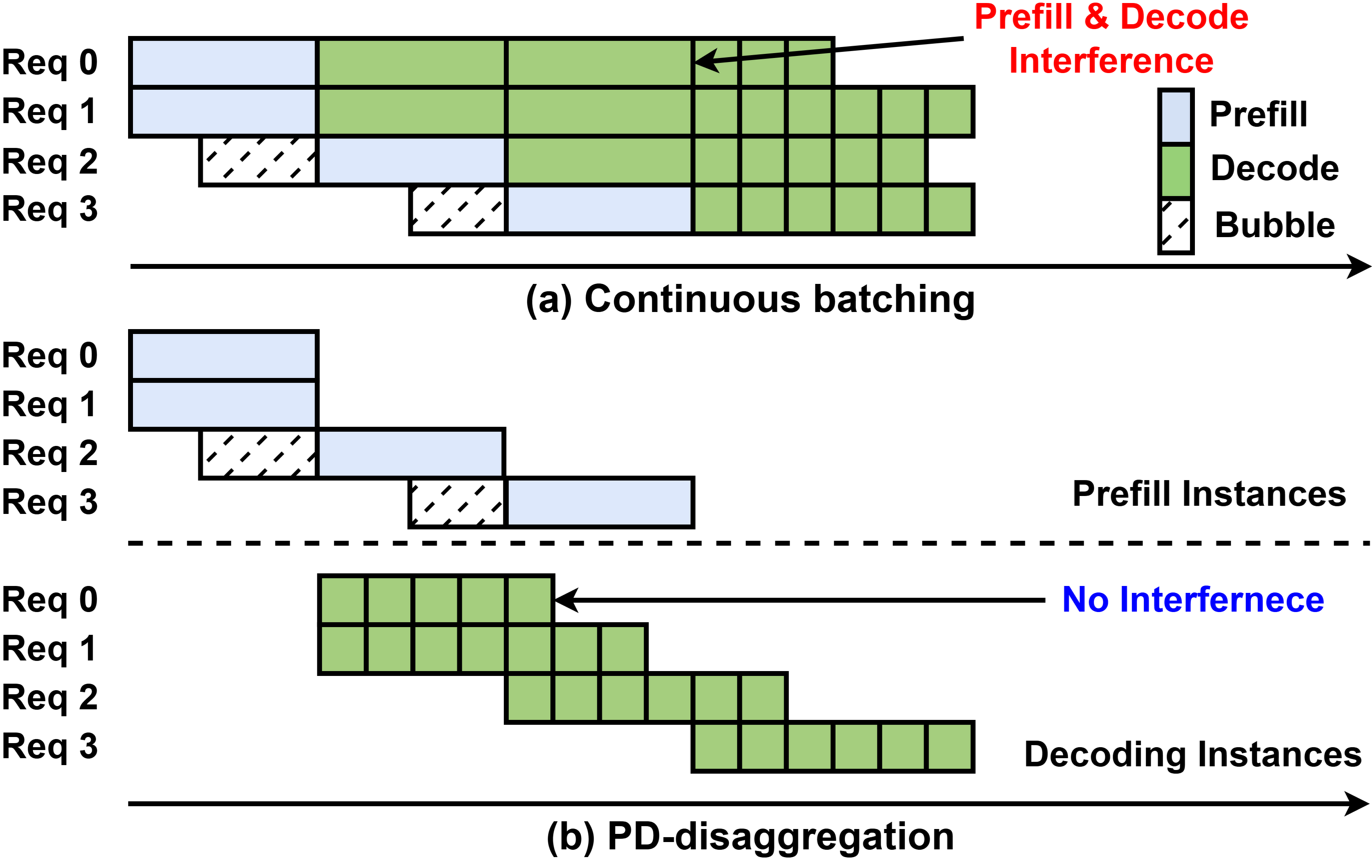}
\vspace{-0.05in}
\caption{Timing diagrams of an LLM serving system.}
\label{fig:batching}
\vspace{-0.2in}
\end{figure}

Batching is essential for improving LLM inference throughput. While naïve batching amortizes weight loading, it becomes inefficient as diverging output sequences shrink the effective batch size. Continuous batching~\cite{orca, sarathi} addresses this by dynamically admitting requests (Fig.~\ref{fig:batching}(a)); however, it introduces severe prefill–decode interference. This serializes short-latency decoding behind long-latency prefill, leading to excessive tail latency in Time Per Output Token (TPOT).
Even with chunked prefill~\cite{sarathi}, this interference remains a major source of latency overhead: small chunks improve TPOT over continuous batching but increase Time To First Token (TTFT) due to repeated KV cache loading for prefill at each chunk, whereas large chunks inflate TPOT as decoding becomes synchronized with the chunk prefill time.
\vspace{-0.1in}
\subsection{Prefill-Decode (PD) Disaggregation}


To eliminate interference between prefill and decoding, PD-Disaggregation~\cite{distserve, windserve, splitwise, dynamo} has been proposed. 
The core idea is to separate two phases by allocating dedicated GPU pools: prefill instances execute compute-intensive operations, while decoding instances handle memory-bound ones. As illustrated in Fig.~\ref{fig:batching}(b), the decoding latency remains independent of concurrent prefill requests.
This decoupling reduces tail latency in TPOT, enabling the system to meet Service Level Objectives (SLOs) compared to continuous batching.
However, PD-disaggregation also introduces several new challenges. 
The KV cache generated from prefill instance is transferred to the decode instance via device-to-device interconnects such as InfiniBand~\cite{infiniband} or NVLink~\cite{NVLink5}. Due to its large size, this transfer incurs significant overhead, motivating techniques such as layer-wise transfer to hide latency~\cite{splitwise, hybe}. Furthermore, these systems suffer from systemic resource imbalances stemming from the rigid partitioning of devices into dedicated prefill and decode clusters (§\ref{resource_util}). 

\vspace{-0.1in}
\subsection{High Bandwidth Memory (HBM)}
HBM is a 3D-stacked DRAM designed to deliver extremely high memory bandwidth. It consists of a base logic die connected to the external processor through a silicon interposer, with multiple DRAM dies stacked vertically using through-silicon vias (TSVs). High bandwidth is achieved through a large number of I/Os connected to the logic die. Although each DRAM die is connected to only a subset of these I/Os, the 3D-stacked structure enables HBM to operate as multiple independent channels. Consequently, all I/O lines are effectively utilized, providing massive aggregate bandwidth within a single memory package.
One of the major changes in the recently introduced HBM4~\cite{HBM4} is that a logic process is used in its base die, replacing the DRAM process used in previous generations. Leveraging this advanced logic node, HBM4 provides 2,048 I/O lines and delivers up to 2 TB/s of bandwidth, doubling that of HBM3~\cite{HBM3}. 
Moreover, the use of a logic process allows the logic die to incorporate additional circuitry, enabling near-memory computing and moving the memory controller from the host to the memory die~\cite{marvell, alphawave_customHBM4_2024, blocksandfiles_skhy_ai_memory_trio_2025, trendforce_skhy_2025_roadmap}.
\section{Motivation and Analysis}

\begin{figure}[t!]
\centering
\includegraphics[width=0.98\columnwidth]{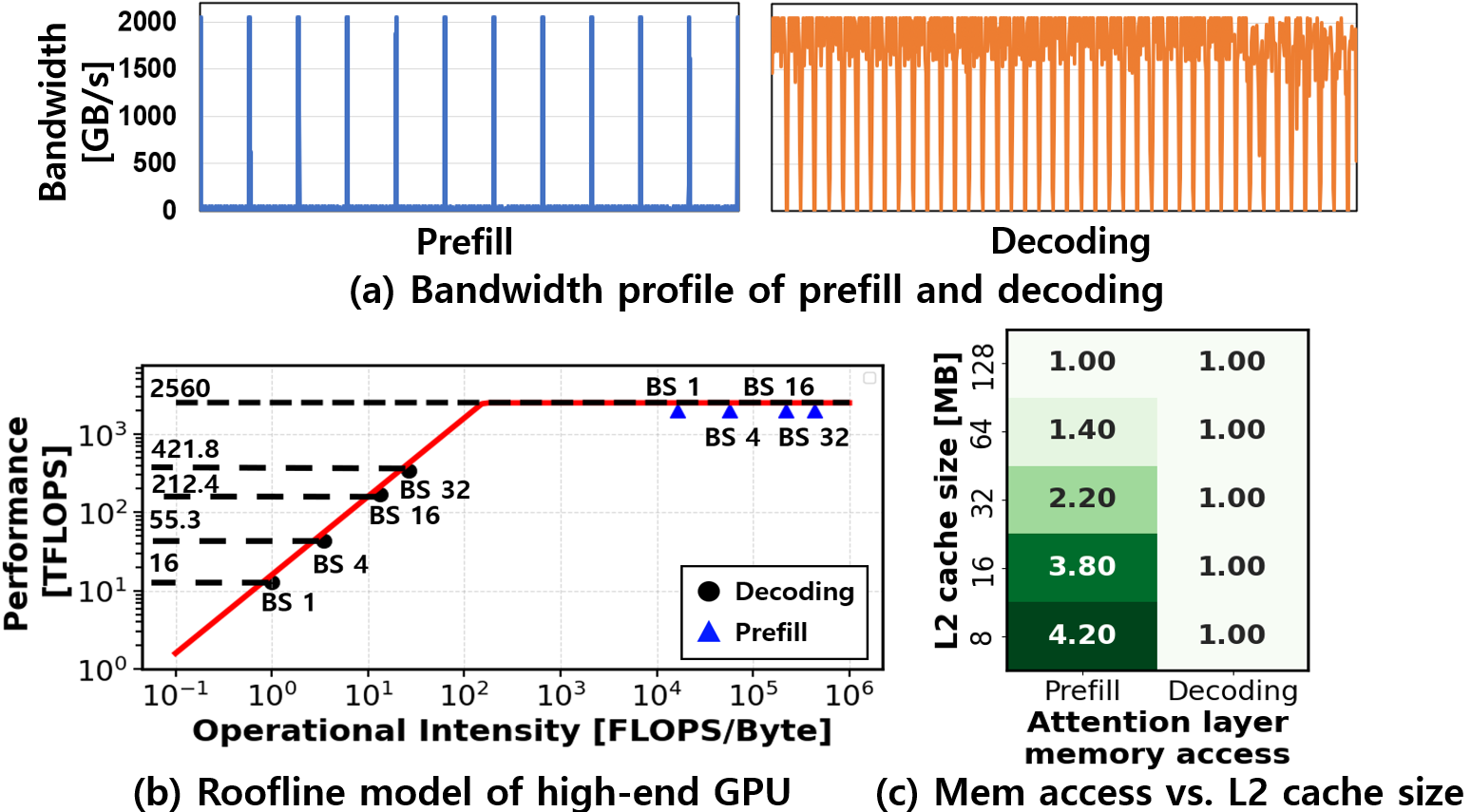}
\vspace{-0.1in}
\caption{Resource under-utilization under GPT3-175B: (a) Bandwidth profile, (b) Roofline model (16K input), (c) Normalized memory access vs. L2 cache size (batch=8, 32K input).}

\label{fig:analysis}
\vspace{-0.2in}
\end{figure}

\subsection{Resource under-utilization in PD-disaggregation}
\label{resource_util}
In PD-disaggregation systems, hardware resources are systematically underutilized~\cite{windserve,hybe}: compute units remain largely idle during decoding, while memory bandwidth remains underused during prefill. Fig.~\ref{fig:analysis}(a) shows that prefill reaches peak bandwidth only briefly, and spends most of its runtime at much lower bandwidth levels, indicating substantial memory underutilization. In contrast, decoding maintains near-peak HBM bandwidth throughout execution.

Roofline analysis (Fig.~\ref{fig:analysis}(b)) on a high-end GPU~\cite{GB300} further shows that decoding requires far fewer compute resources than prefill. Prefill achieves near-peak GPU throughput due to its high arithmetic intensity, whereas decoding is strongly memory-limited. Fig.~\ref{fig:analysis}(c) also shows that decoding benefits little from the GPU’s large L2 cache: reducing L2 capacity by 16$\times$ increases prefill DRAM accesses by up to 4$\times$, indicating strong data reuse, but leaves decoding largely unchanged, implying minimal cache locality.

From these observations, we derive two key insights:



\noindent\circled[black]{1} \textbf{Prefill cannot maintain peak HBM bandwidth.} 
Due to its high arithmetic intensity, prefill exhibits limited memory demand and leaves substantial HBM bandwidth underutilized for most of its execution.

\noindent\circled[black]{2} \textbf{Decoding is compute-light and exhibits minimal cache locality.} 
Decoding utilizes only a small fraction of the GPU’s compute resources and gains little benefit from the L2 cache, making it well suited for a lightweight compute substrate with minimal performance impact.

\begin{figure}[t!]
\centering
\includegraphics[width=\columnwidth]{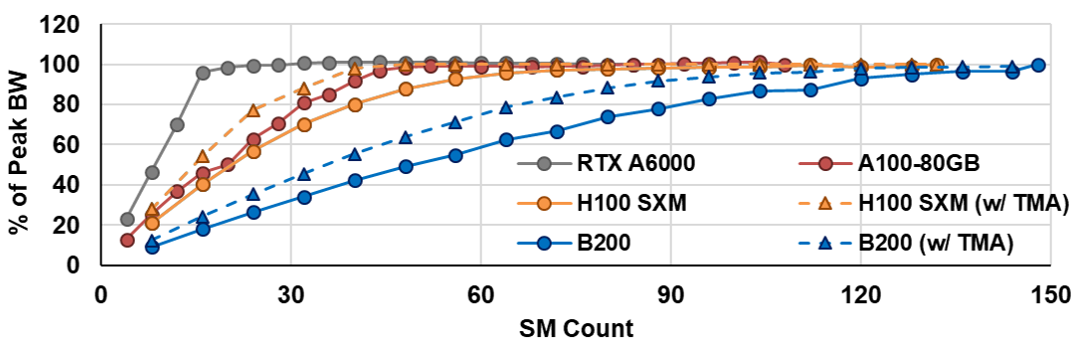}
\caption{Bandwidth analysis of SM partitioning via GreenContext~\cite{greencontexts}, profiled using a streaming read kernel.}
\label{fig:sm_bw}
\vspace{-0.15in}
\end{figure}

\subsection{Limitations of SM-partitioning strategies}\label{sec:SM}
Prior works~\cite{bullet, muxwise} proposed spatial partitioning of streaming multiprocessors (SMs) in GPU to enable single-device PD-disaggregation for improving resource utilization. These methodologies determine an optimal partitioning ratio based on workload characteristics under the Service Level Objective (SLO) targets. While this approach improves resource utilization compared to traditional PD-disaggregation, the inherent SM contention between prefill and decoding remains a significant barrier to maximizing efficiency.

Although decoding is memory-bound and requires limited computation, it still demands a large number of SMs to fully utilize DRAM bandwidth.
This is because each SM supports only a limited number of in-flight memory requests. As a result, using a small number of SMs is insufficient to hide GPU memory latency, including DRAM access, NoC routing and cache hierarchy~\cite{gpumech, MDM, gcom, amali}, and fails to fully utilize the DRAM bandwidth. 

We observed the minimum number of SMs required to saturate memory bandwidth across various GPUs (Fig.~\ref{fig:sm_bw}). The A6000, which provides 768 GB/s via GDDR6, requires only 16 SMs (19\%) to reach 80\% of its peak bandwidth when all SMs are utilized. However, the B200 (with TMA), equipped with 8-stack HBM3e providing 8 TB/s, requires 72 SMs (48.6\%) to achieve the same saturation level.
This allocation for the B200 translates to approximately 1.2 PFLOPS (FP16), which is nearly three times higher than the 422 TFLOPS required for GPT3-175B decoding (at a batch size of 32) as shown in Fig.~\ref{fig:analysis}(b), implying the severe idleness of the computation.
As a result, the contention between SMs makes it nearly impossible to achieve high resource utilization, limiting overall efficiency. 
Recent NVIDIA GPUs introduce the Tensor Memory Accelerator (TMA) to improve bandwidth utilization with fewer allocated SMs. By offloading per-tile address generation and enabling asynchronous bulk data transfers, TMA has been reported to saturate NVLink bandwidth using relatively small number of SMs~\cite{TMA_nvlink}. However, this result does not imply that memory bandwidth utilization is independent of SM allocation. Because the peak HBM bandwidth is substantially higher than the NVLink bandwidth, saturating HBM still requires a much larger number of SMs.
These observations motivate a decoupled architecture.
By offloading decoding to lightweight PNM units embedded in the HBM logic die, memory requests can be generated directly at the source of data, eliminating dependence on SMs and avoiding cache-induced or NoC routing bottlenecks.



\begin{figure*}[t]
\centering
\includegraphics[width=0.96\textwidth]{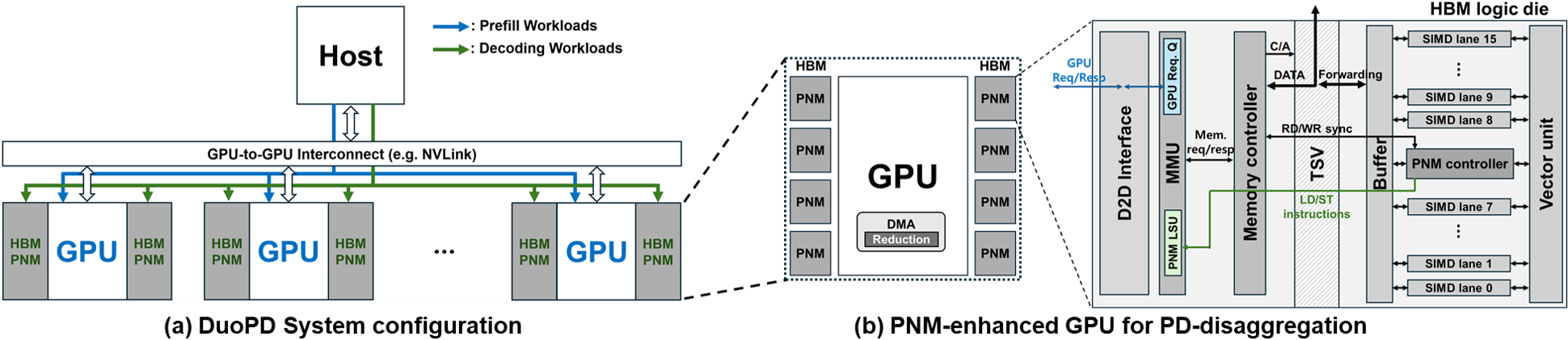}
\vspace{-0.1in}
\caption{Proposed DuoPD system architecture overview.}
\label{fig:Archi}
\vspace{-0.1in}
\end{figure*}

\vspace{-0.1in}
\subsection{Limitations of previous PIM/PNM approaches} \label{sec:approach}

Recent PIM/PNM-based works~\cite{duplex, neupims, attacc} propose heterogeneous systems that couple high-performance GPUs with near-memory compute units. 
These approaches accelerate LLM inference by offloading memory-bound components of decoding (e.g., attention or MoE), and therefore rely on increasing internal memory bandwidth.
However, existing PIM/PNM designs suffer from fundamental limitations. 
On-die PIM approaches~\cite{neupims,attacc,newton,samsung_pim}, which integrate compute logic within DRAM arrays, impose significant overhead on DRAM dies, resulting in severe cell density loss. 

On the other hand, HBM PNMs integrate compute units within the logic die, avoiding this density penalty~\cite{HBM_PNM}.
However, HBM PNMs are typically constrained by the shared TSV-to-interposer interface: the internal bandwidth available to PNM via TSV is the same as the external I/O bandwidth exposed to the GPU in current HBMs.

To overcome this limitation, a recent PNM work, Duplex~\cite{duplex} introduces additional TSVs in the power TSV region to expand internal bandwidth.
However, this approach incurs substantial area and power overheads.
The additional TSVs extend beyond the logic die into the DRAM dies, introducing a 9\% area overhead in DRAM~\cite{duplex}.
This overhead can translate into even larger cell capacity loss due to DRAM peripheral circuit expansion (e.g., HBM-PIM~\cite{samsung_pim} reports a 25\% area overhead leading to a 50\% capacity loss).
Furthermore, Duplex underestimates this overhead by assuming linear scaling with TSV count. In practice, densely integrating TSV I/Os within the power TSV region introduces severe routing contention with existing TSV-to-interposer datapaths, significantly increasing layout complexity.
Moreover, the 4$\times$ bandwidth expansion incurs a substantial power overhead of 100.48 W per stack under HBM4 configurations~\cite{folded}, corresponding to a 2.42$\times$ increase over the baseline of 41.6 W per stack.
Overall, such aggressive bandwidth scaling compromises the feasibility advantages of PNM.

In addition to physical overheads, the increase in bandwidth also introduces a fundamental compute scaling bottleneck. 
To fully utilize the increased memory bandwidth, compute resources must scale proportionally. Otherwise, the system becomes compute-limited.
This constraint significantly restricts the effective operating range of Duplex when offloading the full decoding.
For example, in HBM4 configurations, Duplex is limited to an arithmetic intensity of 4~Op/B, constraining it to small batch sizes (e.g., batch size 4) and limited GQA group sizes.
Scaling compute resources further to match the expanded bandwidth would incur prohibitive area and power costs, making such designs impractical.

Beyond these limitations, prior PIM/PNM approaches are also restricted by their disaggregation granularity.
They typically adopt \textit{intra-decode disaggregation}, where only memory-intensive components of decoding are offloaded to PIM/PNM, while the remaining operations are executed on GPUs within a conventional PD-disaggregation framework.
This partial offloading limits achievable efficiency, as decoding still depends on GPU resources and requires frequent data movement between GPU and PNM, incurring additional overhead.

In contrast, DuoPD adopts a fundamentally different approach along two key dimensions: \textbf{bandwidth management} and \textbf{disaggregation granularity}.
Rather than scaling bandwidth, DuoPD employs a \textbf{bandwidth sharing} approach, as illustrated in 
Fig.~\ref{fig:intro}(b), to ensure hardware feasibility.
Furthermore, instead of intra-decode offloading, DuoPD performs \textbf{inter-phase disaggregation} within a single device, executing prefill on the GPU and the entire decoding phase on PNM over shared memory.
This design significantly improves system-wide resource utilization and performance, offering a practical and novel direction for PNM beyond conventional bandwidth-scaling approaches.

\vspace{-0.1in}
\section{DuoPD: Hybrid GPU-PNM Architecture}
\subsection{System overview}

Unlike conventional PD-disaggregation systems that use separate GPU pools, DuoPD eliminates this partitioning by executing prefill on GPU cores and decoding on lightweight PNM units in the HBM logic die (Fig.~\ref{fig:Archi}). This design is motivated by the limitations of SM partitioning in GPUs: many SMs are required to saturate memory bandwidth, while PNM can achieve full utilization with fewer resources. DuoPD shares TSV I/Os between the GPU and PNM without altering the DRAM core die. This allows PNM to leverage idle bandwidth during the prefill phase, distinguishing it from prior PIM/PNM approaches~\cite{duplex, neupims, attacc} that necessitate increased internal bandwidth. Consequently, it supports asynchronous and concurrent execution of prefill and decoding within a single GPU–PNM engine with minimal interference.


DuoPD is built on three key design techniques. \textbf{First}, we design a compact PNM architecture that supports optimized dataflow for GQA and batching, while enabling full decoding under the area and power constraints. \textbf{Second}, to mitigate data movement conflicts on shared TSV I/O, we propose a direct forwarding dataflow from TSVs to PNM, coordinated by the memory and PNM controllers, along with an Memory Management Unit (MMU) and request arbitration policy to minimize interference between prefill and decoding. \textbf{Finally}, we adopt a hierarchical data mapping scheme optimized for PNM while remaining transparent to GPU execution.

\vspace{-0.1in}
\subsection{Microarchitecture of DuoPD} \label{microarch}
\begin{figure}[t]
\centering
\includegraphics[width=0.98\columnwidth]{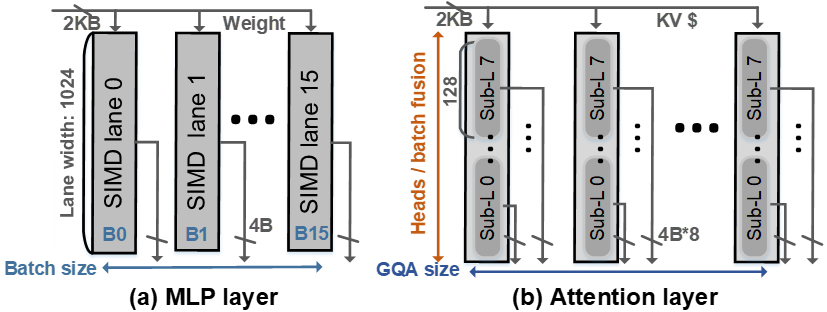}
\vspace{-0.12in}
\caption{Single instruction multiple data (SIMD) unit dataflow for LLM decoding execution.}
\label{fig:SIMD_unit}
\vspace{-0.1in}
\end{figure}



To enable concurrent prefill on GPU cores and decoding on PNM units, we design a DuoPD device as shown in Fig.~\ref{fig:Archi} (b). Following the Blackwell architecture~\cite{GB300}, each GPU is paired with eight HBM4 stacks, and each HBM4 stack has a PNM unit on its logic die.
The die-to-die (D2D) interface bridges the TSVs ascending to the DRAM cores and the interposer routing to the GPU. With HBM4, the transition to an advanced logic process reduces the interface circuitry footprint, enabling the relocation of the memory controller from the GPU to the HBM logic die, which is essential for the centralized arbitration to manage memory access between the PNM units and the GPU~\cite{marvell, alphawave_customHBM4_2024, blocksandfiles_skhy_ai_memory_trio_2025, trendforce_skhy_2025_roadmap} (§\ref{interference}).
DuoPD supports inter-PNM communication (e.g., All-Reduce) using a GPU DMA engine with an in-flight reduction unit (e.g., sum, max), leveraging GPU die area freed by memory controller relocation to avoid impacting compute resources~\cite{conccl,compressing}.


The PNM module comprises the following components:

\textbf{Processing units: }
The SIMD unit contains 16 lanes to support 16 Op/B arithmetic intensity, with each lane operating on 1,024 elements wide and equipped with FP16 multipliers and 8-way adder trees for dot-product operations. At 1 GHz, it delivers a data rate of 2 KB per cycle (2 TB/s), fully matching the bandwidth of an HBM4 stack with 2,048 TSV I/Os.
The SIMD unit is further optimized for MLP and attention layers through a flexible lane architecture (Fig.~\ref{fig:SIMD_unit}). For MLP layers, it processes batched requests by broadcasting weights across lanes for parallel execution, while attention layers exploit a sublane structure, where each lane is divided into eight 128-wide sublanes to enable concurrent processing across heads and requests. With HBM channel interleaving and KV-head broadcasting for GQA, the unit maximizes parallelism under the coordination of the PNM controller.

The vector unit (VU) handles non-MAC operations, such as activation functions and softmax. In particular, it supports online softmax~\cite{online_softmax}, enabling FlashDecoding~\cite{flashdecoding} for efficient attention with sequence parallelism~\cite{helix,sp,loongserve}.
The buffer is sized at 1MB based on performance sweeps, as decoding performance saturates beyond this capacity. It stores inputs and partial results while serving as a dedicated write buffer for PNM requests. Jointly managed by the memory and PNM controllers for TSV data forwarding, the buffer requires 64KB in HBM4, consisting of 32 entries of 32B per pseudo channel across 64 channels, which is well within the 1MB budget.

\textbf{Control units: }
The PNM controller orchestrates all PNM operations. It receives the configuration of decoding workloads and a start signal from the host, and manages the overall decoding process using instructions stored in the instruction buffer. The PNM controller forwards tensor metadata (e.g., base address and size) and load/store instructions to the MMU for SIMD and VU operations.

The MMU efficiently arbitrates concurrent GPU and PNM memory requests to minimize prefill–decode interference. GPU requests are directly enqueued into the GPU request queue, while the PNM load-store unit (LSU) translates dynamic-shape tensor load/store instructions from the PNM controller into 32-B memory requests compatible with the Memory Controller (MC). During this process, the PNM LSU also enforces memory protection checks on the accessed address ranges to ensure system stability between GPU and PNM. 

\begin{figure}[t!]
\centering
\includegraphics[width=0.98\columnwidth]{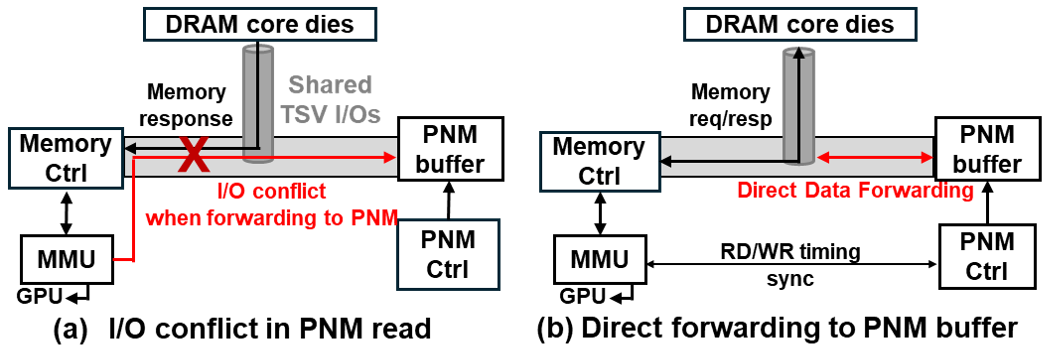}
\vspace{-0.1in}
\caption{(a) I/O conflict in indirect PNM routing for read operations. (b) MC-bypassing direct data forwarding to PNM.}
\label{fig:mc_flow}
\vspace{-0.15in}
\end{figure}

\subsection{GPU-PNM Interference Challenges in DuoPD}
\label{interference}

We introduce solutions to address several challenges of GPU-PNM interference in DuoPD as follows, aiming the minimal performance impact between prefill and decode.

\textbf{Mitigating Shared I/O Confilct. }
As shown in Fig.~\ref{fig:Archi}(b), the D2D interface and MC, connected to the GPU via the interposer, are placed on one side of the TSV array, while PNM units reside on the opposite side due to area and routing constraints. Consequently, PNM requests must traverse the TSV region to reach the shared MC. 
However, due to keep-out zone constraints and dense I/O routing~\cite{KOZ1, KOZ2}, the TSV region offers limited routing resources for high-throughput data movement between the MC and PNM.
To alleviate this limitation, we extend the TSV I/O interface to be shared between the MC and a PNM-side buffer.

However, this shared TSV I/O introduces contention, as the MC and PNM buffer cannot simultaneously utilize the TSV I/Os. For example, Fig.~\ref{fig:mc_flow}(a) illustrates the case when the MC receives data from the DRAM core that corresponds to a PNM request. In this case, the data must be arbitrated by the MMU and then routed back to the PNM buffer, reoccupying the TSV I/O. This interferes with continuous data transfers between the DRAM core and MC, which are critical for achieving peak bandwidth. Importantly, duplicating MCs on both sides of the TSV is not viable, as a shared MC is required to maintain a unified memory space for model parameters across prefill and decode.


To mitigate this contention, we enable direct data forwarding to the PNM units, effectively bypassing the MC for data movement, as illustrated in Fig.~\ref{fig:mc_flow}(b).
In this architecture, data is transferred directly between the DRAM core and the PNM buffer, while the MC maintains global synchronization. For read operations, the MC issues the command and tracks the deterministic timing for the data return; the PNM controller, synchronized with the MC, then latches the data directly from the TSVs. Similarly, for write operations, the MC coordinates the timing by sending a synchronization signal to the PNM controller. This ensures that the PNM buffer drives the write data onto the TSVs in precise alignment with the MC's request to the DRAM core. 

Consequently, our shared TSV I/O interface avoids PNM routing complexity and the MC–PNM interface carries only lightweight control signals, while the heavy data payloads are offloaded to the direct forwarding path, effectively reducing data round trip latency, and data movement energy.

\begin{figure}[t!]
\centering
\includegraphics[width=\columnwidth]{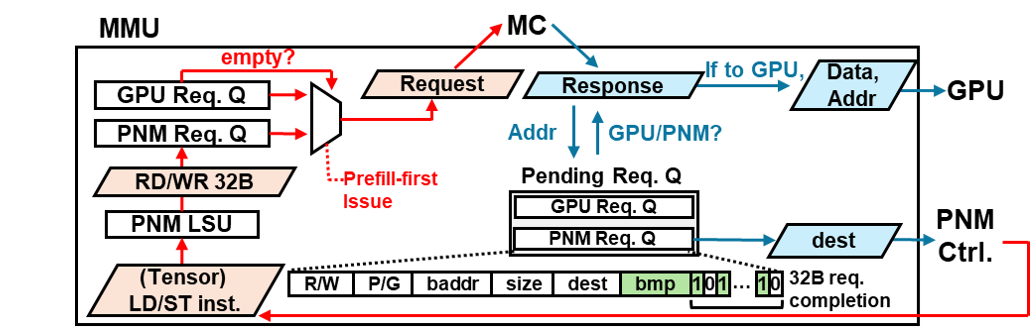}
\caption{Block diagram of the memory management unit for request arbitration and data routing over shared HBM.}
\label{fig:mc_arch}
\end{figure}

\begin{figure}[t!]
\centering
\includegraphics[width=0.98\columnwidth]{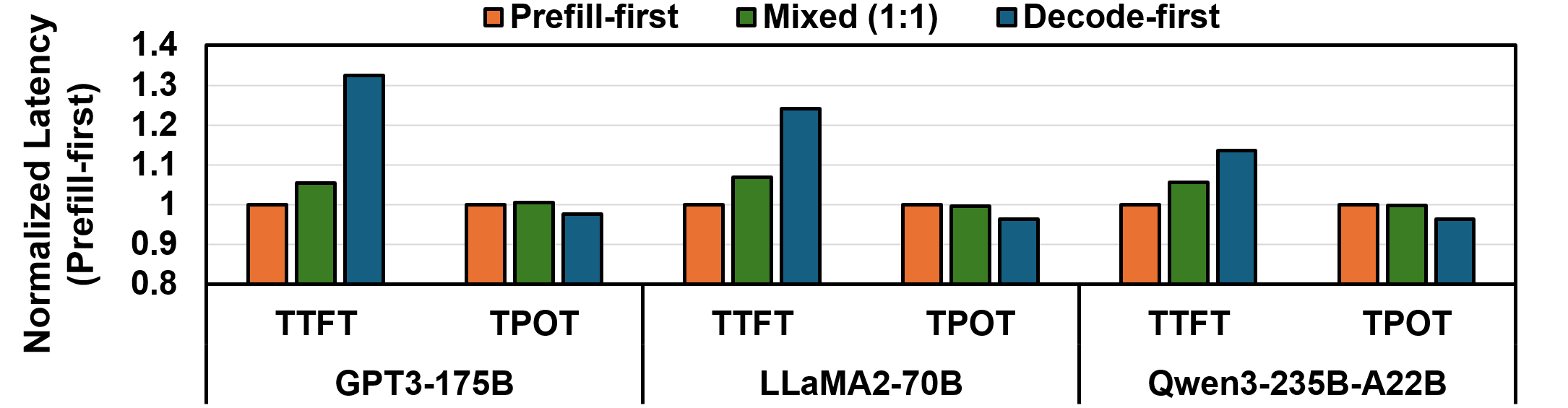}
\vspace{-0.1in}
\caption{Latency across various scheduling policies normalized to the Prefill-first policy.}

\label{fig:latency}
\vspace{-0.2in}
\end{figure}

\textbf{GPU-PNM Memory Arbitration Design. }
Fig.~\ref{fig:mc_arch} shows the organization of MMU that arbitrates memory requests from the GPU and PNM before issuing them to the shared MC. Memory requests generated by the GPU are delivered via the D2D interface and enqueued in the GPU request queue. PNM requests originate from tensor-level load/store instructions issued by the PNM controller, which is decomposed by the PNM LSU into 32-B read/write requests, and inserted into the PNM request queue. The MMU schedules requests from both queues according to a configurable arbitration policy, which directly impacts the Time To First Token (TTFT) and Time Per Output Token (TPOT).

As shown in Fig.\ref{fig:latency}, we evaluate three arbitration policies under prefill–decode interference: \textit{prefill-first}, \textit{decode-first}, and \textit{mixed}. 
The results reveal that TTFT is more sensitive to the arbitration policy than TPOT.  
This is because decoding generates more persistent and bandwidth-intensive memory traffic than prefill. 
Under the decode-first policy, PNM requests dominate the MC transaction queue, causing GPU starvation and significantly increasing TTFT.
In contrast, TPOT shows limited sensitivity to the policy. 
Although bursty prefill requests can temporarily delay decoding, this effect is amortized over the long execution time of decoding due to the low total memory footprint of prefill.
As a result, the average TPOT remains similar across policies.

To further assess the impact of stalls caused by bursty prefill requests, we measure the p99 Time-Between-Tokens (TBT). For GPT3-175B with a batch size of 16 and a context length of 16K, the p50 TBT is 15.71 ms, while the p99 TBT increases only slightly to 17.18 ms. This limited impact arises because the stalls typically affect only one or two layers within each decode step, while the dozens of remaining layers proceed uninterrupted, effectively amortizing the overhead over the entire step. Moreover, when measuring TPOT, the stall overhead is further amortized across all generated tokens.

Overall, the prefill-first policy provides the best end-to-end performance. 
While prioritizing prefill slightly increases TPOT for models with high decode bandwidth demand (e.g., 
LLaMA and Qwen), this overhead is outweighed by the reduction in TTFT. 
Moreover, even under the prefill-first policy, DuoPD achieves higher decoding performance than conventional PD-disaggregation, as it allocates more effective memory bandwidth to decoding under the same GPU budget.

After servicing requests at the MC, memory responses are re-arbitrated by the MMU and routed to their destination. The MMU matches each response address with entries in the pending request queues to identify whether it corresponds to a GPU or PNM request. 
To mitigate the in-flight request bottlenecks, PNM request entries are created at tensor granularity and include the request type, target (GPU/PNM), tensor base address, access size, destination address in the PNM buffer, and a bitmap tracking completion of constituent 32-B requests. The bitmap is updated upon each sub-request completion, and the entry is retired once all bits are set, triggering a completion signal to the PNM controller. 
For GPU requests, both data and address are returned to the GPU. 
For PNM requests, the data is directly forwarded to the PNM buffer upon arrival, and only the resolved destination address is sent to the PNM controller. 
This design eliminates redundant data movement through the memory controller and minimizes prefill–decode interference, enabling efficient concurrent execution of GPU and PNM workloads.

\vspace{-0.03in}
\section{Data Mapping Techniques for DuoPD} \label{Model-Parallelism}
Model parallelism addresses how to efficiently distribute parameters and KV cache across multiple GPU nodes during LLM inference serving. Commonly used techniques include Tensor Parallelism (TP)~\cite{alpa,parallelism}, which partitions weights and merges partial results via all-reduce, Pipeline Parallelism (PP)~\cite{gpipe, pipedream}, which splits layers across devices but suffers from low batching efficiency, and Expert parallelism (EP)~\cite{switch_transformer, gshard, deepspeed} in Mixture-of-Experts (MoE) models, which distributes experts across devices, with tokens dynamically routed to a subset of experts for scalable execution.

Unlike traditional designs, DuoPD experiences a hierarchical execution environment:
\textit{GPU-level parallelism} across GPU instances, and
\textit{PNM-level parallelism} within a single GPU-HBM engine.
We explore this two-tier structure considering system efficiency with prefill-decode concurrency.

Additionally, we discuss memory space management for PNM and GPU to enable concurrent access to a shared address space, where model parameters are shared while KV caches are exclusively allocated and accessed by each device.


\subsection{GPU-Level Model Parallelism} \label{GPU-MP}
In DuoPD, GPU-level model parallelism differs from conventional PD-disaggregation in a key aspect:
GPU-level parallelism must be identical for both prefill and decoding, because the two phases share the same model parameters. Mismatched partitioning would make parameter sharing infeasible.
Prefill prefers PP due to its high compute intensity, while decoding prefers TP because batch sizes are small. However, with high-bandwidth interconnects such as NVLink, TP communication overhead is modest, whereas PP significantly degrades decoding throughput. Therefore, DuoPD adopts \textbf{TP over PP} at the GPU level. For MoE models, we further incorporate EP at the GPU level, following prior systems.
\subsection{PNM-Level Model Parallelism}
\begin{figure}[t!]
\centerline{\includegraphics[width=0.98\columnwidth]{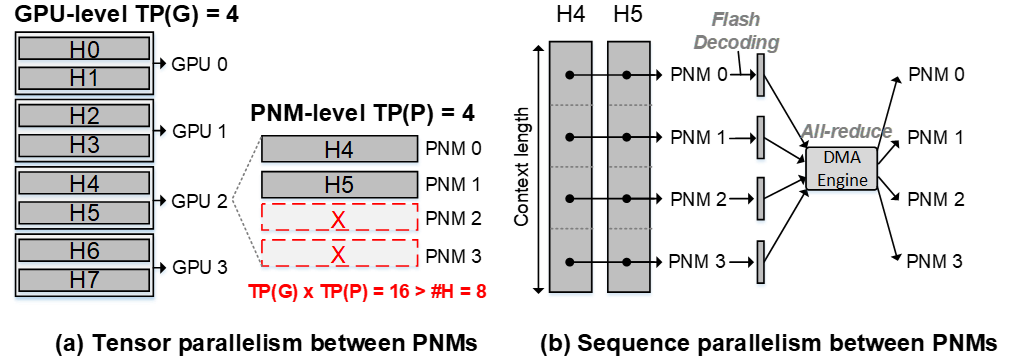}}
\caption{Model parallelism for inter-PNM model mapping.
}
\label{fig:model_parallelism}
\end{figure}

Parallelizing decoding across PNMs is fundamentally constrained by both architectural and algorithmic factors. Conventional PP is ineffective in this setting because distributing transformer layers across PNMs serializes prefill accesses, activating only one HBM stack at a time and severely underutilizing memory bandwidth. Also, applying EP at the PNM level is inefficient, as dynamic expert selection creates load imbalance across PNMs, leaving some HBM-PNM stacks idle while the most heavily loaded PNM determines the overall decoding latency.
TP, while effective for most layers, faces a critical limitation in the attention stage. Since GPU-level TP is already required for high decoding throughput, applying additional TP across PNMs multiplies the overall parallelism degree, which quickly becomes too large to support attention head partitioning (Fig.~\ref{fig:model_parallelism}(a)).
This leads to impractical mappings with fractional heads per device, particularly for models with a limited number of KV heads (e.g., GQA). 

To overcome these limitations, we adopt sequence parallelism (SP)~\cite{helix,sp,loongserve} that partitions the KV cache along the sequence dimension rather than across attention heads, as illustrated in Fig.~\ref{fig:model_parallelism}(b). This avoids the scalability bottlenecks in TP while enabling balanced utilization of PNM resources. Sequence partitioning introduces a challenge for softmax normalization, which requires reduction across the full sequence. To address this, we leverage FlashDecoding~\cite{flashdecoding}, which performs attention in a tiled manner with numerically stable partial reductions. Each PNM computes attention locally while maintaining intermediate statistics, followed by a lightweight inter-PNM all-reduce to complete normalization.

Inter-PNM communication in DuoPD such as SP is confined within a single GPU/HBM package and therefore requires neither GPU-GPU synchronization nor communication over external interconnects such as NVLink. Even assuming a conservative GPU NoC latency of 400--500 ns based on~\cite{uncovering_GPU_noc}, the resulting overhead is only $\sim$2\% of the $>$24 $\mu$s single-layer decode time. Thus, inter-PNM communication introduces only a minor latency overhead.

Overall, this approach enables scalable and efficient PNM-level parallelism without excessive TP scaling or bandwidth underutilization, allowing DuoPD to effectively exploit aggregate HBM bandwidth across PNMs while remaining compatible with GPU-level parallelism.

\begin{figure*}[t!]
\centerline{\includegraphics[width=\textwidth]{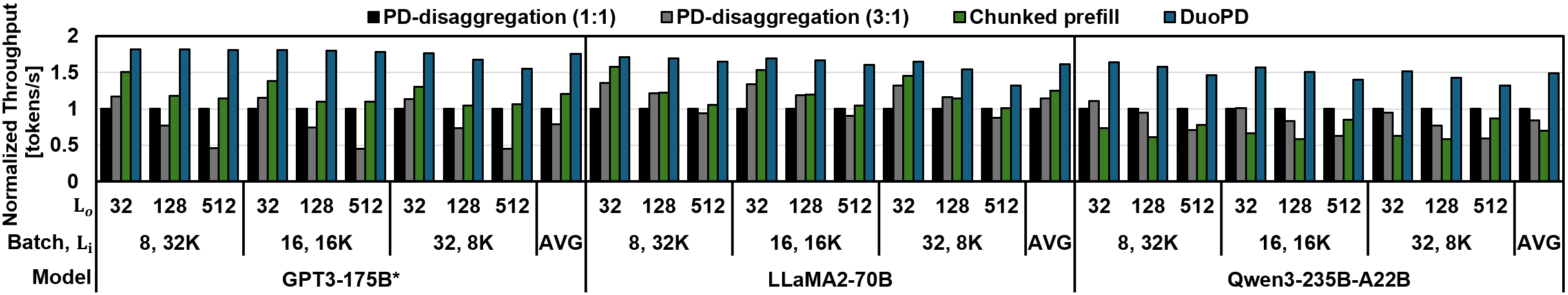}}
\caption{The normalized end-to-end throughput results. All baselines use 8 GPUs. $L_i$:input length, $L_o$:output length. For GPT3-175B, the batch size for PD-disaggregation (3:1) is reduced due to capacity constraints.}
\label{fig:e2e_throughput}
\end{figure*}

\subsection{Memory Space Management}
Because the GPU and PNM share a unified memory space, DuoPD enforces access protection for LLM inference data. Model parameters are read-only during inference and can therefore be shared by both GPU and PNM without coherence mechanisms. In contrast, KV cache and intermediate values are mutable, but they are accessed exclusively by either the GPU or the PNM during their respective execution phases and are not concurrently shared for read-write updates. Thus, DuoPD avoids GPU-PNM coherence overheads, requiring only a prefill-to-decode handoff that makes GPU-written KV cache visible to PNM before decode begins.

On the PNM side, the lightweight MMU and LSU enforce the read-only policy for shared model parameters. Before issuing a request to the shared memory controller, the PNM LSU validates that accesses to the parameter address space are read-only. This localizes protection to the PNM path while enabling safe concurrent execution with the GPU.

\section{Evaluation}
\subsection{Evaluation setup}

\textbf{Baselines:}
We evaluate two types of baseline systems: PD-disaggregation, where prefill and decoding are executed on separate GPUs, and chunked prefill, where both phases run within the same GPU. For a fair comparison, we assume that all systems are configured with eight GPU chips, each equipped with eight HBM4 stacks, providing 2 TB/s bandwidth~\cite{HBM4_BW} and 48 GB capacity~\cite{HBM4} per stack.
The GPU cores provide 2.5 PFLOPS in FP16, consistent with GB300~\cite{GB300_spec}.

In the PD-disaggregation setup, we assume that the prefill and decoding instances are connected through NVLink5~\cite{NVLink5}, which provides 900 GB/s bandwidth for KV cache transfer. We evaluate two configurations: a 1:1 ratio between prefill and decoding instances, 
and a 3:1 ratio, which allocates more GPUs to the prefill side. 
In the chunked prefill setup, we use a chunk size of 1K for dense models, and 4K for MoE models, to avoid excessive degradation in TPOT and TTFT.

\textbf{PNM Configuration:}
We configure the PNM in DuoPD using SIMD-style processing elements (PEs). All PNM units operate at 1 GHz, and each SIMD lane consists of 1,024 FP16 multipliers with 8-way adder trees. Each lane delivers a throughput of 2 TFLOPS, and with 16 lanes operating in parallel, the total peak performance reaches 32 TFLOPS. The PNM is also equipped with a 1 MB on-chip SRAM buffer. 

\textbf{Simulation:}
To evaluate performance, we built a cycle-level simulator\footnote{https://github.com/SNU-VLSI-DRAM/PD-intra-disaggregation-with-HBM-NDP} that emulates both GPU and PNM operations. The simulator was integrated with DRAMSim3~\cite{dramsim3} to accurately capture prefill-decode memory interference. The simulator receives user configurations (e.g., TP, PP, EP, hardware specifications), allocates parameter and KV cache addresses to HBMs, and generates inference requests. For each request, separate prefill and decoding instructions are generated for the GPU and PNM. The GPU scheduler and PNM controller modules execute these instructions independently. 


\textbf{Workload:}
We evaluate three large-scale models in a server-level environment: GPT3-175B~\cite{GPT} (dense, Multi-Head Attention (MHA)), LLaMA2-70B~\cite{LLaMA2-70B} (dense, Grouped-Query Attention (GQA)), and Qwen3-235B-A22B~\cite{qwen3} (Mixture-of-Experts (MoE), GQA).
We focus on long-context workloads, where PD-disaggregation is most effective~\cite{beyond_the_buzz}. Table~\ref{tab:dataset} shows the average input and output lengths of modern long-context datasets~\cite{longbench, loogle}. Based on these results, we use input lengths of 8K–32K tokens and output lengths of 32–512 tokens, covering various scenarios.

\begin{table}[t]
  \centering
  \caption{Average lengths of long-context datasets~\cite{longbench, loogle}.} \vspace{-0.06in}
  \label{tab:dataset}
  \fontsize{9}{11}\selectfont
  \begin{tabular}{c|c}
    \toprule
    \makecell{\textbf{Category}} & \makecell{\textbf{(Avg input, Avg output)}} \\
    \midrule
    Question\&Answer & (7.1K, 4) / (7.3K, 17) / (22K, 14) \\
    Summarization & (10K, 706) / (21K, 254) \\
    Code completion & (6.0K, 22) / (21K, 24) \\
    \bottomrule
  \end{tabular}
  \vspace{-0.1in}
\end{table}



\begin{figure*}[t!]
\centerline{\includegraphics[width=\textwidth]{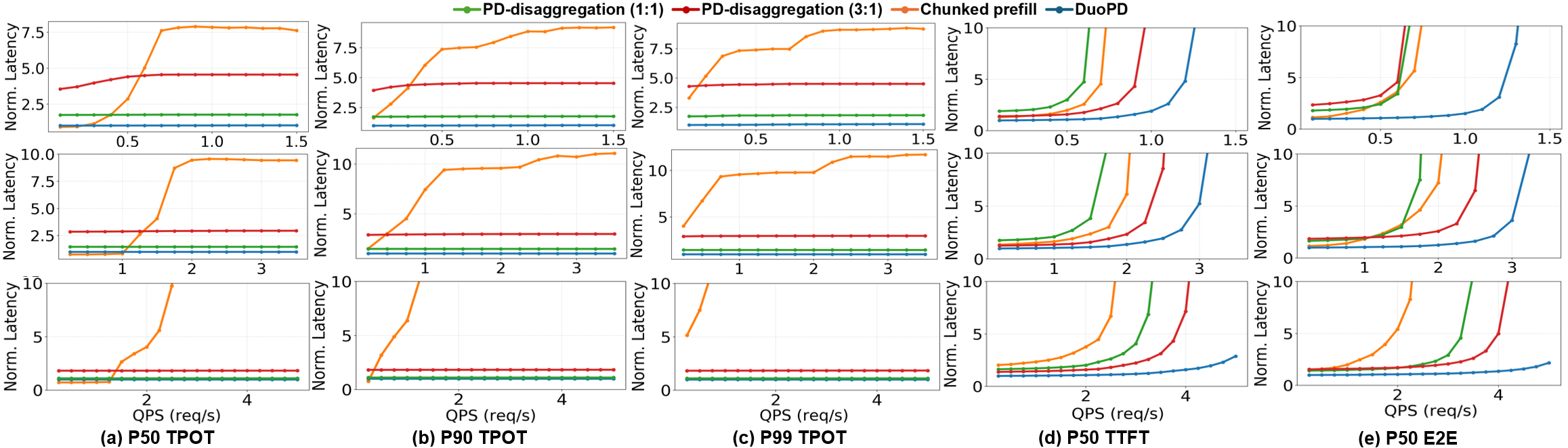} 
}
\caption{Normalized latency of GPT3-175B (top), LLaMA2-70B (middle), and Qwen3-235B-A22B (bottom) across varying QPS. The maximum decode batch size is bounded by the memory capacity limit. Input length is 32K and output length is 128.}
\label{fig:qps_results}
\vspace{-0.06in}
\end{figure*}

\subsection{Throughput}

Fig.~\ref{fig:e2e_throughput} reports normalized end-to-end throughput. DuoPD consistently outperforms all baselines, achieving average improvements of 1.46–2.23$\times$ on GPT3-175B, 1.29–1.61$\times$ on LLaMA2-70B, and 1.45-1.78$\times$ on Qwen3-235B-A22B.
In particular, DuoPD achieves up to 1.76$\times$ and 2.23$\times$ higher throughput than 1:1 and 3:1 PD-disaggregation, respectively.
This improvement stems from the DuoPD architecture, which fundamentally changes resource allocation across phases. 
In baseline PD-disaggregation systems, GPUs and their attached HBMs are partitioned between prefill and decoding (e.g., 4:4 or 6:2), thereby limiting prefill to only a fraction of GPU resources and decoding to a fraction of available HBM bandwidth. 
As a result, prefill cannot fully utilize GPU compute, while decoding is constrained by reduced memory bandwidth. 
In chunked prefill systems, prefill and decoding are executed on the same GPU, but their conflicting resource demands introduce contention for compute, cache, and memory bandwidth, resulting in degraded throughput. 
In contrast, DuoPD assigns all GPU compute resources to prefill while offloading decoding entirely to PNM units that exploit otherwise underutilized HBM bandwidth, with minimal cross-phase interference. 
This separation enables simultaneous and efficient utilization of compute and memory resources, resulting in significantly higher end-to-end throughput.
For long outputs at batch size 32, the gain decreases as PNM is optimized for batch size 16, which leads to suboptimal decoding performance. Nonetheless, DuoPD consistently achieves the 1.31-3.44$\times$ higher performance than baselines.

Additionally, even in prefill-heavy cases ($L_o$ = 32), DuoPD achieves 1.25–1.60$\times$ and 1.08–2.42$\times$ higher throughput than 3:1 PD-disaggregation and chunked prefill, respectively, despite their larger prefill compute pools. Note that chunked prefill is suboptimal for MoE models due to the overhead of repeatedly loading experts. For long output lengths ($L_o$ = 512), DuoPD achieves 1.31–3.92$\times$ higher throughput than both 3:1 PD-disaggregation and chunked prefill.


\subsection{TPOT, TTFT, E2E latency}

Fig.~\ref{fig:qps_results} shows TPOT, TTFT, and End-To-End (E2E) latency results measured under varying queries per second (QPS) for each model. To emulate a dynamic serving environment, requests arrive according to a Poisson distribution based on the target QPS, and the request lengths follow a Gaussian distribution.
Across all evaluated cases, DuoPD consistently achieves lower latency than the other baselines and maintains stable latency at higher QPS.



DuoPD reduces TPOT by up to 42.61\% and 77.66\% compared to 1:1 and 3:1 PD-disaggregation, respectively. Chunked prefill experiences excessive tail latency as QPS increases due to prefill–decode interference. 
DuoPD also achieves lower TTFT by utilizing more GPUs for prefill, and this advantage becomes more pronounced as QPS increases. As TTFT grows rapidly under high QPS due to queuing delays, DuoPD can maintain stable TTFT at higher QPS. We use the TTFT at the lowest QPS as a no-queuing baseline and measure the QPS at which TTFT increases by 3$\times$, which serves as a proxy for the onset of significant queuing delays and service-level degradation. At this point, DuoPD supports up to 2.13$\times$, 1.39$\times$, and 3.12$\times$ higher QPS than 1:1 PD-disaggregation, 3:1 PD-disaggregation, and chunked prefill, respectively, with similar trends observed for E2E latency.

Overall, these results show that DuoPD maintains stable performance under dynamic request arrivals and can achieve higher QPS, resulting in improved serving goodput.

\subsection{Comparison with PIM/PNM Decode Accelerator}

As discussed in Section~\ref{sec:approach}, prior PIM/PNM-based approaches fundamentally differ from DuoPD in that they aim to increase the internal bandwidth and offload a subset of decode operations (e.g. attention and MoE) to PIM/PNM. 
To compare with these approaches, we implement bank-level parallelism PIM~\cite{neupims, attacc} and Duplex~\cite{duplex}, a representative HBM-PNM accelerator, within a PD-disaggregation system. For PIM, we assume 50\% effective memory capacity to account for area overhead that reduces cell density~\cite{samsung_pim}, and offload decode attention to PIM. For PNM, we implement 1:1 PD-disaggregation (1:1 PD-D) on Duplex by offloading decode attention and MoE, assuming no capacity loss due to unquantified area overhead. To match Duplex’s bandwidth, we further evaluate DuoPD with 4$\times$ higher PNM bandwidth (DuoPD (4$\times$ BW)) and configurations with 4$\times$ more compute resources, including Duplex (1:1 PD-D, 4$\times$ compute) and DuoPD (4$\times$ BW + 4$\times$ compute).

\begin{figure}[t!]
\centerline{\includegraphics[width=\columnwidth]{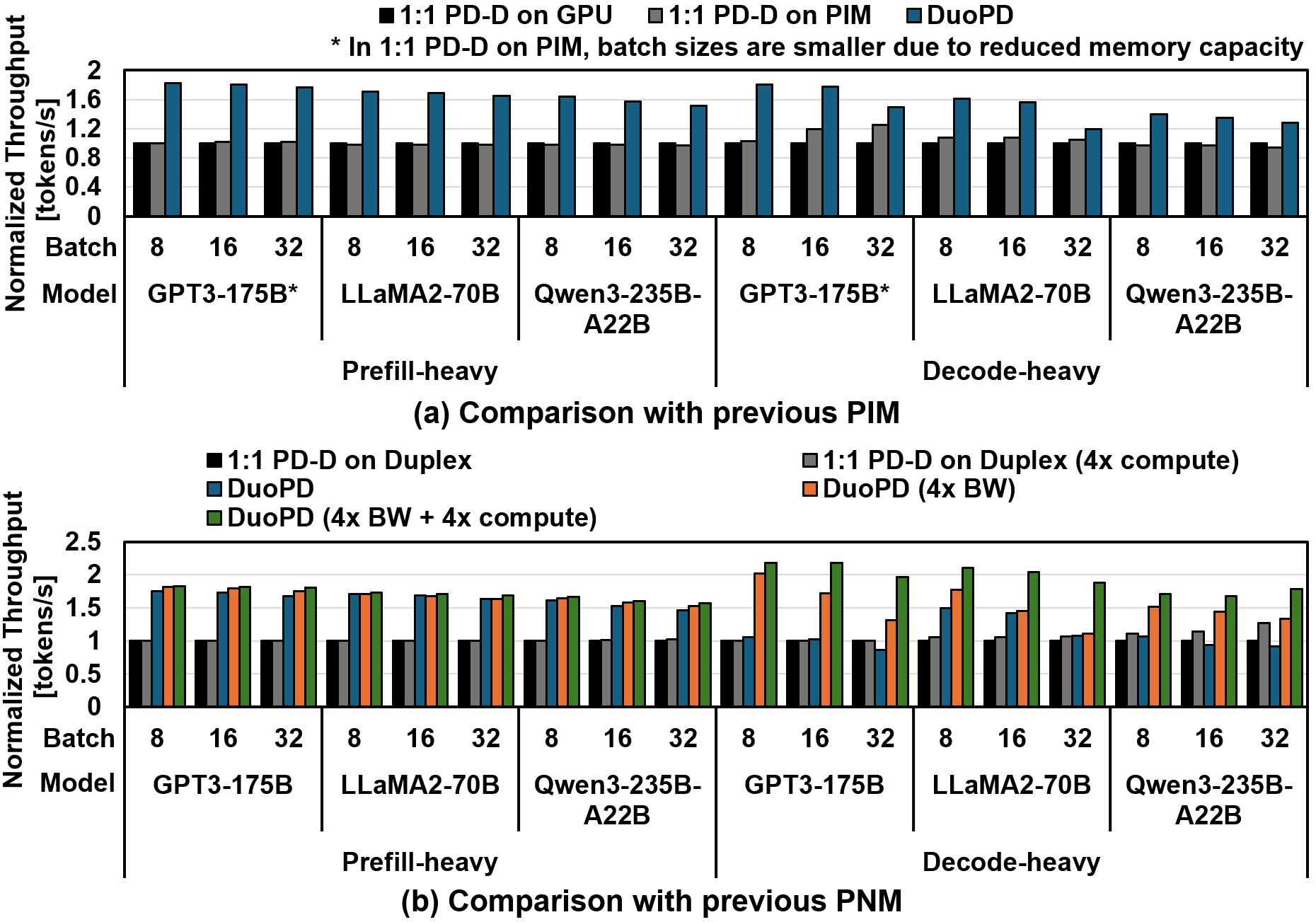}}
\caption{Normalized end-to-end throughput of PIM/PNM and DuoPD: Input length: 32K/16K/8K (batch size 8/16/32), Output length: 32 (prefill-heavy), 1K (decode-heavy).}
\label{fig:PIM_PNM_comparison}
\vspace{-0.2in}
\end{figure}

Fig.~\ref{fig:PIM_PNM_comparison} (a) and (b) show throughput comparisons of DuoPD with PIM- and PNM-based accelerators, respectively, across three LLMs under prefill- and decode-heavy workloads.

In \textbf{Prefill-heavy} workloads, accelerating decode with PIM/PNM has little impact on overall throughput. As a result, DuoPD achieves up to 1.82$\times$ higher throughput than 1:1 PD-D on both PIM and Duplex.



In \textbf{Decode-heavy} workloads, PIM improves throughput by accelerating decode attention, achieving up to 1.26× higher throughput than 1:1 PD-D on GPU for GPT3-175B (Fig.~\ref{fig:PIM_PNM_comparison} (a)). However, DuoPD consistently outperforms 1:1 PD-D on PIM, achieving up to 1.80$\times$ higher throughput due to the high arithmetic intensity of GQA and reduced PIM capacity.

Fig.~\ref{fig:PIM_PNM_comparison} (b) shows that DuoPD achieves up to 1.49$\times$ higher throughput than PNM for non-MoE models. This result highlights the limitation of bandwidth scaling, which requires additional compute units to fully utilize the increased bandwidth. This limitation is also observed in DuoPD (4$\times$ BW), where throughput decreases with increasing batch size and in GQA-based models. For MoE models, although Duplex benefits from the lower arithmetic intensity of expert computation, the performance gain compared to DuoPD is modest (1.09$\times$) and insufficient to justify its substantial area and power overhead of increased bandwidth.
Even with increased compute resources, Duplex provides limited benefits despite incurring significant area and power overheads, as it offloads only a subset of decoding components (e.g., attention and MoE experts). In contrast, with identical bandwidth and compute scaling, DuoPD (4$\times$ BW + 4$\times$ compute) achieves the best performance by executing full decoding while concurrently performing prefill on the GPU cores, delivering up to 2.18$\times$ higher throughput than 1:1 PD-D on Duplex.


These results demonstrate that phase-level disaggregation of prefill and decoding across the GPU and PNM, as enabled by DuoPD, is more effective than relying on aggressive PIM/PNM bandwidth scaling to accelerate only a subset of decoding operations.

\subsection{Comparison with SM partitioning}

\begin{figure}[t!]
\centerline{\includegraphics[width=\columnwidth]{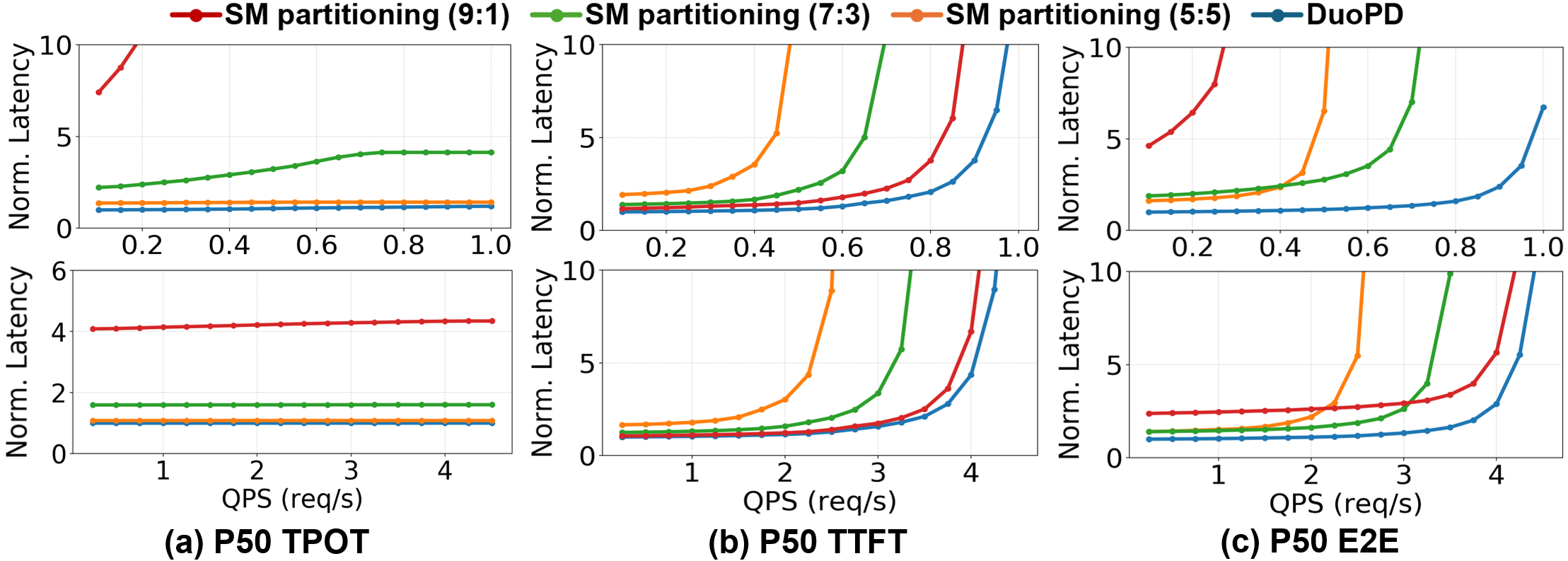} \vspace{-0.02in}}
\caption{Normalized latency of GPT3-175B (top) and Qwen3-235B-A22B (bottom) across varying QPS. Input length is 32K and output length is 128.}
\label{fig:SM_QPS}
\vspace{-0.06in}
\end{figure}

We also compare DuoPD with intra-GPU PD-disaggregation via SM partitioning.
To this end, we extend our simulator using the measured relationship between memory bandwidth and the number of SMs on the B200, as characterized in §\ref{sec:SM}.
We evaluate three SM partitioning configurations (prefill:decode= 9:1, 7:3, and 5:5) using GPT3-175B and Qwen3-235B-A22B, measuring p50 TPOT, TTFT, and E2E latency.
As shown in Fig.~\ref{fig:SM_QPS}, DuoPD consistently achieves lower latency than SM partitioning across all metrics.
SM partitioning exhibits an inherent trade-off between TPOT and TTFT: allocating more SMs to decoding improves TPOT but degrades TTFT, and vice versa. 
For example, a 5:5 partition achieves TPOT comparable to DuoPD, but increases TTFT by 1.93$\times$ and 1.66$\times$ even at the lowest QPS. 
Conversely, a 9:1 partition matches DuoPD's TTFT, but incurs over 5$\times$ higher TPOT.
This trade-off arises because prefill and decoding share the same GPU resources in SM partitioning, leading to contention for both compute and memory bandwidth. 
Even with dynamic SM partitioning schemes~\cite{bullet,muxwise}, such contention cannot be fundamentally eliminated, as both phases compete for shared on-chip and off-chip resources. 
In contrast, DuoPD separates compute-bound prefill and memory-bound decoding across GPU and PNM, respectively, reducing cross-phase contention and enabling efficient simultaneous utilization of compute and memory resources.

\subsection{Resource Utilization Enhancement in DuoPD}

\begin{figure}[t!]
\centering
\includegraphics[width=\columnwidth]{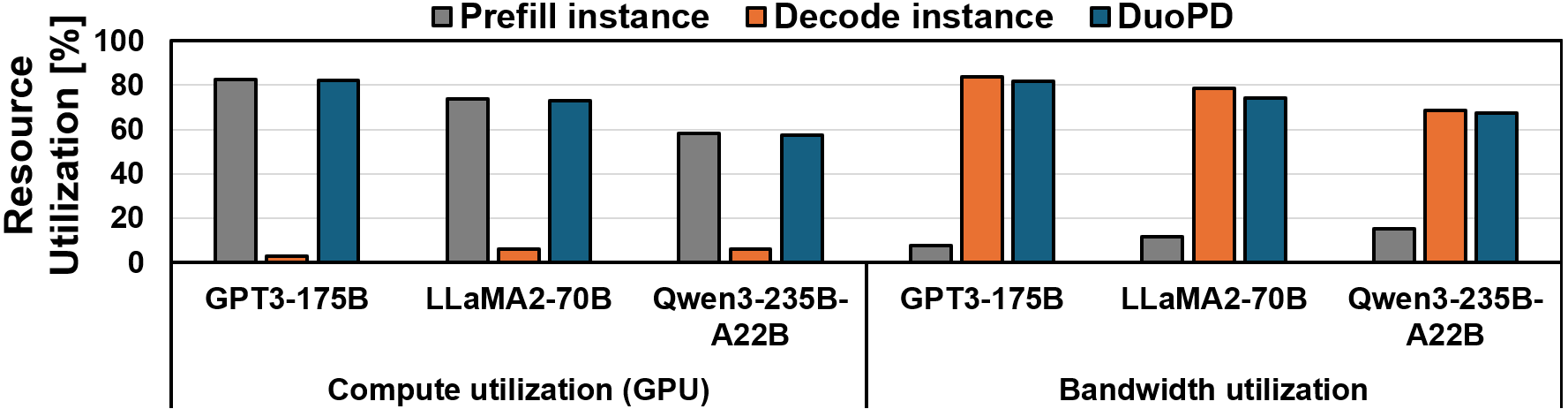}
\caption{GPU compute and memory bandwidth utilization for batch size 16 and input length 16K.}
\label{fig:resource_util}
\end{figure}
Fig.~\ref{fig:resource_util} shows the GPU core compute utilization and memory bandwidth utilization for each model.
DuoPD executes prefill on GPU cores while simultaneously performing decoding on PNM. As a result, it achieves GPU compute utilization comparable to prefill instances ($<$1\% difference), while providing 9.50$\times$–30.62$\times$ higher utilization than decode instances.
Similarly, DuoPD maintains memory bandwidth utilization within 5\% difference to decode instances, which is up to 10.60x higher than prefill instances. 
These results validate the key idea of DuoPD: by sharing memory and enabling asynchronous execution of prefill and decoding across GPU and PNM, DuoPD effectively maximizes both GPU compute utilization and HBM bandwidth utilization.

\subsection{Area and Power analysis} \label{area}


To evaluate the area feasibility of integrating the PNM unit into the HBM logic die, we use the estimated logic die area based on die photographs of HBM4~\cite{hbm4_samsung}, which is approximately 140.8~mm\textsuperscript{2}. Out of the area, 38.6~mm\textsuperscript{2} remains available after excluding TSV and PHY regions. We synthesize all logic components of the PNM using Synopsys Design Compiler in a 28~nm CMOS process and estimate SRAM area via CACTI~\cite{cacti}. All results are then scaled to 7~nm following standard scaling rules~\cite{scaling}. The total PNM area is 6.51~mm\textsuperscript{2}, which agrees with published 7~nm SRAM macro densities~\cite{sram} and is consistent with prior PNM implementations such as Duplex~\cite{duplex}. The SIMD unit accounts for the largest area of 3.64~mm\textsuperscript{2}, and PNM buffer and Vector unit occupies 1.25~mm\textsuperscript{2} and 1.56~mm\textsuperscript{2}, respectively. This suggests that the proposed PNM comfortably fits within the available logic die budget of future HBM generations such as HBM4. Moreover, it does not require modifications to the DRAM core die, thereby avoiding cell density loss.

\begin{figure}[t!]
\centerline{\includegraphics[width=\columnwidth]{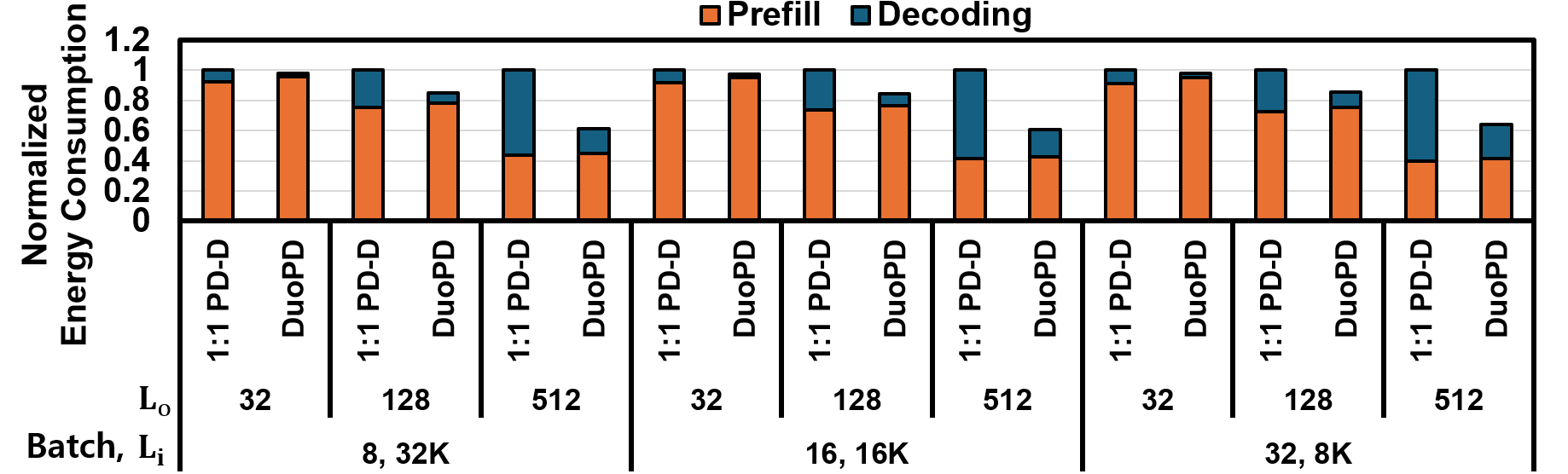} }
\caption{Normalized end-to-end energy consumption of GPT3-175B between 1:1 PD-disaggregation and DuoPD.}
\label{fig:energy}
\vspace{-0.12in}
\end{figure}

We also estimate power from the synthesized PNM components in a 28~nm CMOS technology and scale them to 7~nm node similar to area analysis.
The total PNM power per HBM stack is 6.45 W, dominated by the SIMD unit (4.9 W), which are similar values to that of previously reported compute units~\cite{lincoln,sigma,monde}.
To evaluate the thermal impact of the PNM units, we conduct simulations using HotSpot~\cite{Hotspot}. We model a server-grade cooling solution based on a vapor-chamber heat sink, consistent with prior works~\cite{3D-path, stratum}. We use an effective convection resistance of 0.0146 K/W~\cite{vapor-chamber} and configure a 16-high HBM stack. Each DRAM core die dissipates 2 W~\cite{power}, and the PNM units are distributed within the 38.6~mm\textsuperscript{2} available region of the logic die. Our thermal model also accounts for the heat generated by the GPU core die and its thermal coupling with the HBM stack. Under these conditions, at an ambient temperature of 45°C, the simulated peak temperature is approximately 76°C, below the 85°C refresh-safety threshold~\cite{3D-path}. DuoPD also imposes a smaller thermal burden than prior PNM designs such as Stratum and Duplex: it consumes less PNM power than Stratum's reported 45 W~\cite{stratum}, and, unlike Duplex~\cite{duplex}, DuoPD avoids additional TSV I/O power.

To compare the system-level energy efficiency of DuoPD against 1:1 PD-disaggregation, we model GPU and memory power using publicly reported values. We use the GB300’s reported 1400 W power consumption~\cite{GB300_spec}, and set the per-stack HBM4 power to 41.6 W following previous works~\cite{folded}. NVLink communication energy is modeled using 8 pJ/b~\cite{bandwidth,GTC}. As DuoPD avoids round-trip data movement through the silicon interposer during decoding, we conservatively apply a 7\% reduction in HBM4 energy~\cite{folded}.

Fig.~\ref{fig:energy} reports the normalized end-to-end energy consumption for GPT3-175B. Prefill energy remains similar across both systems. During decoding, however, DuoPD consumes on average only 32.4\% of the GPU’s decoding energy due to the significantly lower power of PNM execution and the elimination of NVLink KV-cache transfers. As a result, the total system-level energy savings grow with output length: DuoPD reduces end-to-end energy by 2.1\% for 32-token outputs, 14.8\% for 128 tokens, and up to 37.9\% for 512 tokens.

\section{Related Works}
Prior work has explored heterogeneous deployments for prefill and decoding in PD-disaggregation systems, leveraging the distinct compute and memory characteristics of the two phases. Splitwise~\cite{splitwise} exploits GPU heterogeneity by deploying NVIDIA A100s for memory-bound decoding and H100s for compute-bound prefill. Hybe~\cite{hybe} offloads decoding to a custom NPU with low compute density to alleviate the compute underutilization of decoding GPUs. However, Hybe does not address the bandwidth underutilization of prefill instances and incurs higher cost than DuoPD because it requires separate memory systems built from expensive HBM devices.
More recently, NVIDIA announced Rubin CPX~\cite{rubin}, designed to accelerate long-context inference workloads. In this architecture, prefill is executed on CPX while decoding is executed on HBM-equipped Rubin GPUs, and CPX employs GDDR7~\cite{gddr7} instead of HBM to reduce the memory cost of prefill instance. While heterogeneous deployment with the Rubin-CPX alleviates memory bandwidth utilization inefficiencies, it still suffers from the compute under-utilization of GPUs dedicated to decoding and require separate DRAMs for prefill and decode instances.

\section{Conclusion}
In this work, we presented DuoPD, a hybrid GPU–PNM architecture that addresses the resource underutilization problem in conventional PD-disaggregation systems through inter-phase disaggregation. Unlike prior PIM/PNM approaches that rely on increasing internal bandwidth, DuoPD improves resource utilization by leveraging bandwidth sharing, exploiting the idle bandwidth of prefill phase through concurrent execution of GPU-prefill and PNM-decoding. To realize this design, we introduced a compact PNM microarchitecture, an MMU design to manage shared TSV I/O, and a hierarchical data mapping scheme for efficient GPU–PNM execution. Our simulation results show that DuoPD achieves up to 2.23$\times$ higher throughput and 77.66\% lower TPOT than state-of-the-art PD-disaggregation systems, demonstrating a practical and cost-efficient path toward high-throughput LLM serving.

\bibliography{main}

\end{document}